\documentclass[prl,aps,showpacs,twocolumn,superscriptaddress]{revtex4}
\usepackage{graphicx}
\usepackage{amssymb}
\usepackage{amsmath}
\usepackage{color}

\newcommand{\beq}{\begin{eqnarray}}
\newcommand{\eeq}{\end{eqnarray}}

\usepackage{psfrag}

\begin{document}

\title{Entanglement in Four-Dimensional SU($3$) Gauge Theory}

\author{Etsuko~Itou}
\email{eitou@post.kek.jp}
\affiliation{KEK Theory Center, High Energy Accelerator Research Organisation, Tsukuba
305-0801, Japan}

\author{Keitaro~Nagata}
\email{knagata@post.kek.jp}
\affiliation{High Energy Accelerator Research Organisation (KEK), Tsukuba
305-0801, Japan}

\author{Yoshiyuki~Nakagawa}
\affiliation{RIISE,
Hiroshima University, Higashi-Hiroshima, Hiroshima, 739-8521, Japan}

\author{Atsushi Nakamura}
\email{ atsushi@rcnp.osaka- u.ac.jp}
\affiliation{RIISE,
Hiroshima University, Higashi-Hiroshima, Hiroshima, 739-8521, Japan}
\affiliation{Theoretical Research Division, Nishina Center, RIKEN, Wako 351-0198, Japan}
\affiliation{Research Center for Nuclear Physics (RCNP), Osaka University, Ibaraki, Osaka, 567-0047, Japan}
\affiliation{School of Biomedicine, Far Eastern Federal University, Sukhanova 8, Vladivostok 690950, Russia}

\author{V.I. Zakharov}
\email{vzakharov@itep.ru}
\affiliation{School of Biomedicine, Far Eastern Federal University, Sukhanova 8, Vladivostok 690950, Russia}
\affiliation{ITEP, B. Cheremushkinskaya 25, Moscow, 117218 Russia}
\affiliation{Moscow Inst. Phys. \& Technol, Dolgoprudny, Moscow Region, 141700 Russia.}

\begin{abstract}
We investigate the quantum entanglement entropy for the four-dimensional Euclidean SU(3) gauge theory. 
We present the first non-perturbative calculation of the entropic $c$-function ($C(l)$) of SU(3) 
gauge theory in lattice Monte Carlo simulation using the replica method.
For $0 \leqslant l \leqslant 0.7$~fm, where $l$ is the length of the subspace, 
the entropic $c$-function is almost constant, indicating conformally invariant dynamics.
The value of the constant agrees with that perturbatively obtained from free gluons, 
with 20 \% discrepancy. 
When $l$ is close to the Hadronic scale, the entropic $c$-function decreases smoothly, and it is consistent with zero within error bars
at $l \gtrsim 0.9$ fm. 
\end{abstract}

\maketitle

Quantum entanglement is a fascinating phenomenon that was first 
highlighted by the Einstein-Podolsky-Rosen paradox~\cite{Einstein:1935rr} and has remained a focus
of research activity for decades.  
If there is a system  in a pure quantum state, 
measurements on  a subsystem $A$ determine the results of measurements
on its complement $B$, even if no causal communication is possible
between the two measurements.  
The entanglement entropy 
$S_{A}$ of subsystem $A$ is defined as von 
Neumann entropy corresponding to the reduced density matrix $\rho_{A}$:
\begin{equation}\label{sa}
S_{A}~=~-\mathrm{Tr}_{{\cal H}_A} \rho_A \log (\rho_A), 
\end{equation} 
where $\rho_A = \mathrm{Tr}_{{\cal H}_B} \left[ \rho_{\rm tot}  \right]$, and
it is assumed that the total Hilbert space is a direct product of
two subspaces corresponding to the subsystems considered, 
$\mathcal{H}_{\rm tot}=\mathcal{H}_{A}\otimes \mathcal{H}_{B}$.

More generally, studies of the entanglement entropy become central 
in cases of complex systems with strong interactions, where the properties of the ground state
cannot be evaluated directly. 
In particular, the notion of quantum entanglement is crucial
for the theory of quantum phase transitions, i.e., non-thermal phase
transitions at temperature $T=0$~\cite{Sachdev,Vidal:2002rm,Calabrese:2004eu}.
In physics of black holes, consideration of the quantum entanglement is central to
discussions of the information paradox \cite{Hawking:1974sw}, which challenges the consistency of general relativity and
quantum mechanics.

Applications to field theory are more recent. First of all,  
the entanglement entropy is ultraviolet divergent in field theory
\cite{Srednicki:1993im}. In more detail, one 
considers the vacuum state
and defines the subsystem $A$ as a slab  of length $l$ in one of
the spatial dimensions, at a fixed time slice. 
Then, the entanglement entropy contains, as its most divergent term,
a term that is proportional to $|\partial A|/a^{d-1}$, 
where $d$ is the number of spatial dimensions, $a$ is the lattice spacing, and $|\partial A|$ 
is the area of the boundary surface between the slab and the rest of the space.
To eliminate this divergence, which depends on details of the UV cut-off, one focuses
on the entropic $c$-function~\cite{Nishioka:2006gr}:
\begin{equation}
C(l)=\frac{ l^3}{|\partial A|}\frac{ \partial S_A}{ \partial l}~,\label{eq:extropic-c-4d}
\end{equation}
where we choose $d=3$. 
$C(l)$ is a finite quantity even in the $a\rightarrow 0$ limit, and it becomes constant as a function of $l$ in the conformal case.

In the present work, we non-perturbatively obtain the entropic $c$-function of SU($3$) gauge theory, 
which describes the dynamics of gluons and has a confinement property.
Although no analytic proof exists yet, 
accumulated numerical evidences imply that quantum chromodynamics (QCD) has a finite mass gap.
At zero temperature, we expect from the asymptotic freedom that the entropic $c$-function 
is approximated by the contribution of non-interacting gluons at short distances $l$. 
On the other hand, at the hadronic scale, $l \sim \Lambda_{\rm QCD}^{-1}$, 
the entropic $c$-function captures the physics of confinement, or strong interactions. 
No analytical calculation of $C(l)$ seems possible in this region. 
In the limit $l\gg \Lambda_{\rm QCD}^{-1}$,
the effective degrees of freedom responsible
for the entanglement apparently reduce to non-interacting glueballs.
It has been suggested~\cite{Klebanov:2007ws} that the entropic $c$-function 
estimated by the correlation function of glueballs shows a Hagedron-type divergence.
Furthermore, several works based on holographic and geometrical approaches found that 
$S_{A}$ undergoes a quantum phase transition at $l_{\rm cr}\sim \Lambda_{\rm QCD}^{-1}$~\cite{Ryu:2006bv,Nishioka:2009un,Ryu:2006ef,Nishioka:2006gr,Klebanov:2007ws,Kol:2014nqa,Witten:1998zw,Lewkowycz:2012mw,Sakai:2004cn}. 
At this critical value of $l$, the entropic $c$-function in the large $N_c$ limit changes its behavior  
from $S_{A}\sim N_c^2$ at short distance
to $S_{A}\sim N_c^0$ at long distance.
Here $N_c$ is the number of colors.

There is a subtlety concerning the
local gauge invariance of the entanglement entropy and of
the Hilbert space for the subspace $A$ on the lattice. 
Although several predictions and definitions for the entanglement entropy for the lattice 
gauge theory have been proposed
~\cite{Kabat:1995eq,Ryu:2006ef, Chen:2015kfa,Ghosh:2015iwa,Soni:2015yga,Aoki:2015bsa,Casini:2013rba,Radicevic:2014kqa,Donnelly:2011hn,Donnelly:2014gva,Velytsky:2008rs,Buividovich:2008gq,Buividovich:2008kq,Nakagawa:2009jk,Nakagawa:2011su}, 
it turns out that some 
definitions give different values for the entanglement entropy.
Recently, a definition that emphasizes the maximally gauge invariance has been proposed in 
Refs.~\cite{Ghosh:2015iwa,Aoki:2015bsa}.
The entanglement entropy in the replica method~\cite{Buividovich:2008gq,Buividovich:2008kq,Nakagawa:2009jk, Nakagawa:2011su} 
agrees with  
that of the maximal gauge-invariant definition.
In our work, we utilize the replica method following Refs.~\cite{Buividovich:2008gq,Buividovich:2008kq, Nakagawa:2009jk,Nakagawa:2011su},  
and we obtain the entropic $c$-function numerically.

The replica method is a powerful technique for calculating the entanglement entropy.
Based on this method, the entanglement entropy is given by the following equation:
\beq
S_A 
&=& \lim_{n \rightarrow 1} \left[ -\frac{\partial}{\partial n} \ln \left( \mathrm{Tr} \rho_A^n  \right)  \right] .
\eeq
Here, $n$ is an integer, and it is referred to as a replica number.
The period in the temporal direction for field variables in subsystem $A$ is $n$ times as long as that of subsystem $B$.
The trace of the $n$-th power of the reduced density matrix $\rho_A$ is given by the ratio of the partition functions:
\beq
\mathrm{Tr} \rho_A^n = Z(l,n)/Z^n.
\eeq
Here, $Z$ is the original partition function for the whole system, and
$Z(l,n)$ is the partition function for the system with an $n$-sheeted Riemann surface. 
The subsystem $B$ is a patch of the $n$-th Riemann surface while the subsystem $A$, whose length of one direction is $l$, is defined on a single Riemann surface.

\begin{figure}[h]
\begin{center}
\includegraphics[scale=0.5]{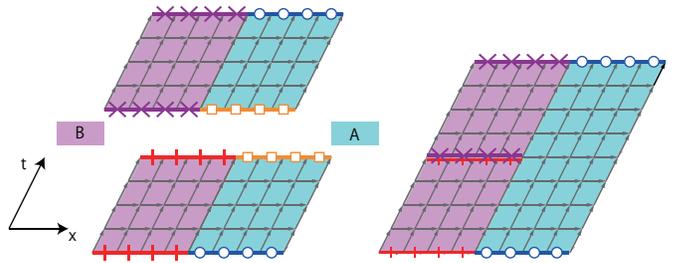}
\caption{Replica lattice and boundary conditions. }
\label{fig:replica-lattice}
\end{center}
\end{figure}

The whole system is realized on a four-dimensional lattice of size $N_s^3 \times N_t$, with lattice spacing $a$, 
where $N_s$ and $N_t$ are the spatial and temporal lattice sizes, respectively.
The system is divided into two subsystems, $A$ and $B$, in the $x$-direction, and the numbers of sites in $x$-direction of $A$ and $B$ are $L$ and $N_s-L$, respectively.
We adopt periodic boundary conditions for all directions.
As explained above, the period of the temporal direction depends on the $x$ coordinate.
We show an example of the boundary condition on the replica lattice with $N_t=4, n=2$, in Fig.~\ref{fig:replica-lattice}.
In the figure, the boundaries with the same symbols in the $t$-direction are matched with each other via the periodic boundary condition.
Thus, in subsystem $B$, the period of the temporal direction for the link variables is $N_t (=4)$ in Fig.~\ref{fig:replica-lattice}, while in $A$, it becomes $(n \cdot N_t)$.
The boundary surface between $A$ and $B$ is extended in the $y$-$z$ plane, and the area is given by $|\partial A|=(N_s a)^2$ in physical unit.

The entropic $c$-function is obtained as the derivative of $\mathrm{Tr} \rho_A^n$ with 
respect to $l$ and $n$.
These derivatives are approximated by
finite differentials with $(\Delta L)=1$ and $(\Delta n)=1$.
We introduce the interpolating action~\cite{Endrodi:2007tq,Fodor:2007sy} given by
\beq
{\cal S}_{\mathrm{int}}= (1- \alpha) {\cal S}_L [U] + \alpha {\cal S}_{L+\Delta L} [U],\label{eq:interpolating-action}
\eeq
where ${\cal S}_L$ and ${\cal S}_{L+\Delta L}$ represent the averaged action density on the replica lattices in which $L$ and $L+\Delta L$
are the lengths of the subsystem $A$.
Here, $U$ denotes the link variables, which are related to SU($3$) gauge fields as $U_{\mu} (\vec{x},t)=\exp(ig_0 A_\mu (\vec{x},t))$ with a bare coupling constant $g_0$.
Now, we can rewrite Eq.~(\ref{eq:extropic-c-4d}) as
\beq
C(l)&=& \frac{  L^3}{ N_s^2 \Delta L} \int_0^1 d\alpha \langle {\cal S}_{L+\Delta L} [U] - {\cal S}_L [U] \rangle_\alpha,
\eeq
where $\langle \cdot \rangle_\alpha$ refers to the Monte Carlo average with the interpolating action ${\cal S}_{\mathrm{int}}$ at a fixed value of $\alpha$.
We regard $l$ in $C(l)$ as $(L+\Delta L/2)a $.

The strategy for calculating the entropic $c$-function using the lattice Monte Carlo simulation consists of five steps:

{\bf Step 1: }
Generate gauge configurations on the replica lattice using Monte Carlo simulation.
The interpolating action ${\cal S}_{\mathrm{int}}$ is used as a weight for the probability.

{\bf Step 2:}
Measure $ {\cal S}_{L+\Delta L} -{\cal S}_{L} $ on each generated gauge configuration, and take an ensemble average of this for each value of $\alpha$.

{\bf Step 3:}
Numerically integrate $\langle {\cal S}_{L+\Delta L} -{\cal S}_{L} \rangle_{\alpha}$  as a function of $\alpha$.

{\bf Step 4:}
Take the continuum limit.

{\bf Step 5:}
Estimate the replica number dependence.\\

We utilize the standard Wilson plaquette action as an action, 
which has one coupling constant, namely the lattice bare coupling constant $\beta=6/g_0^2$.
Gauge configurations are generated by using the pseudo-heatbath algorithm. 
Thus, the link variables ($U$) are updated using the local
 interpolating gauge action ${\cal S}_{\mathrm{int}}$ at a fixed value of $\alpha$.

The details of simulation are as follows.
The simulations were performed with a replica lattice volume $N_s^3 \times N_t =16^4$ and $n=2$.
The simulation parameters and the number of generated configurations are summarized in Table~\ref{table:number-conf-zeroT}.
\begin{table}[h]
\begin{center}
\begin{tabular}{ccccccc}
\hline \hline 
$\beta$&$a$ [fm] & $L=2$    & $L=3$ & $L=4$  & $L=5$  & $L=6$ \\  
\hline
$5.700$ & $0.1707$  &   $12,000$     &  $16,000$  & $72,000$   &  $30,000$ & \\
$5.720$ & $0.1628$ &                    &         & $60,000$   &   & \\
$5.740$  & $0.1555$ &                     &         & $60,000$   &   & \\
 $5.750$ & $0.1520$ &    $12,000$    &   $16,000$ & $67,000$   &  $67,000$ & \\
 $5.770$ & $0.1454$ &                     &         & $30,000$   &   & \\
 $5.780$ & $0.1423$ &                     &         & $30,000$   &   & \\
$5.800$  & $0.1363$ &   $12,000$     &   $16,000$ & $30,000$   & $84,000$  & $52,000$ \\
$5.870$ &  $0.1182$ &           $12,000$     &         &         &        &  \\
\hline \hline 
\end{tabular}
\caption{ Simulation parameters and the number of configurations for ($N_s,N_t, n$) $=$ ($16,16,2$) lattice.}  \label{table:number-conf-zeroT}
\end{center}
\end{table}
Each configuration is separated by $100$ sweeps to avoid the autocorrelation.
We also show the value of the lattice bare coupling constant ($\beta$) and the corresponding length of the lattice spacing ($a$) in physical unit.
The pure Yang-Mills gauge theory is an asymptotically free theory, and it has only one physical scale, $\Lambda_{\mathrm{QCD}}$.
Once we fix a relation between a physical reference scale and a quantity in lattice unit, then all physical quantities can be obtained in physical unit.
We use the relation between the lattice bare coupling constant and the lattice spacing given in Eq.(2.18) of Ref.~\cite{Guagnelli:1998ud}.
Here, as a reference scale,
we utilize the Sommer scale, in which the dimensionless static quark--antiquark force satisfies $r^2 F(r)|_{r=r_0}=1.65$.
To convert a quantity in lattice unit into physical unit, we assume that $r_0=0.5$ fm.

The error is estimated using the bootstrap method. 
Firstly we calculate $\partial S_{A}/ \partial l$ for each bootstrap sample, and then estimate the statistical error from its distribution. 
The typical number of bootstrap samples constructed is $O(10^3 - 10^5)$.

In {\bf Step 2} and {\bf Step 3}, 
for each $L$, we take $11$ points of $\alpha$ between $\alpha=0.0$ to $\alpha=1.0$, at intervals of $\Delta \alpha=0.1$.
The numerical integration is carried out using the cubic polynomial function.
We also numerically investigated the dependence of $C(l)$ on the number of $\alpha$ and the integration formula, 
and found that such effects are sufficiently smaller than the statistical uncertainty.

\begin{figure}[t]
\begin{center}
\hspace{5cm}\includegraphics[scale=0.7]{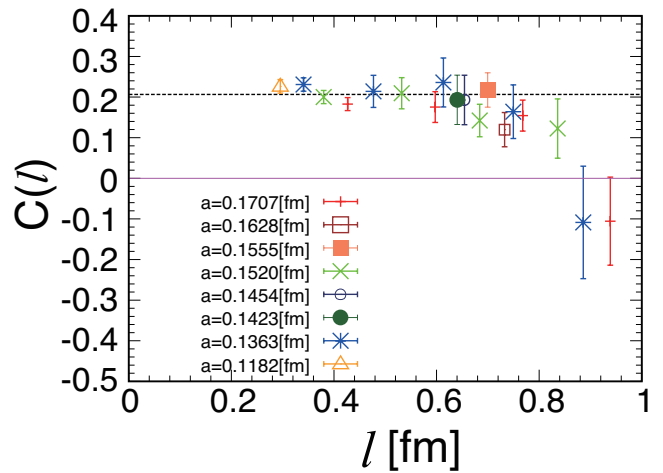}
\vspace{15pt}
\caption{Entropic $C$-function obtained by ($N_s,N_t, n$) $=$ ($16,16,2$) replica lattices. 
The dotted (black) line shows the best-fit value of $C=2.06$.}
\label{fig:C-fn}
\end{center}
\end{figure}

Figure~\ref{fig:C-fn} shows the result for the entropic $c$-function of the pure Yang-Mills theory at zero temperature.
We found that the $c$-function is almost constant in the small $l$ region ($l \lesssim 0.7$ fm);
we fit the data with a constant for the data in the region  $0 \le l \le 0.7$, and obtain the best-fit value
\beq
C=0.206 \pm 0.007,\label{eq:result-Cfn}
\eeq
where the chi-square of degrees of freedom is $0.88$.
Here the error bars denote $1\sigma$ statistical error.
To estimate the systematic uncertainty, we change the range of fitting to $0 \le l \le 0.6$ and $0 \le l \le 0.8$, 
and obtained $C=0.208(8)$ and $C=0.202(7)$, respectively.
The systematic uncertainty of $C$ coming from the choice of the fit range is smaller than the statistical error.

The data shows continuous decrease in the middle $l$ regime, and becomes consistent with zero beyond $l=0.88$ fm. 
The critical temperature of the pure SU($3$) Yang-Mills theory determined by the center 
symmetry breaking is $T_c=280$ MeV, that is, $1/T_c=0.714$ fm \cite{Iwasaki:1996sn}.
The $\Lambda$ scale obtained from the running coupling constant based on the lattice 
simulation is $r_0 \Lambda_{\overline{\mathrm{MS}}}=0.602(48)$~
\cite{Guagnelli:1998ud} and $r_0 \Lambda_{\overline{\mathrm{MS}}}=0.613(2)(25)$~\cite{Gockeler:2004ad}. 
They correspond to $\Lambda_{\overline{\mathrm{MS}}}^{-1}\sim0.831$ fm and 
$\Lambda_{\overline{\mathrm{MS}}}^{-1} \sim 0.816$ fm, respectively,  when we set the 
Sommer scale $r_0=0.5$ fm.
The length of $l$, for  which the $c$-function starts decreasing, is in approximate agreement with these scales.

Next, let us compare our results with those found by other studies.
A numerical simulation for the pure SU($2$) gauge theory was carried out 
in Ref.~\cite{Buividovich:2008kq}.
In SU($2$) gauge theory, the entropic $c$-function shows a clear discontinuity around $l=0.5$ fm,
and it shows an enhancement when the length is slightly less than this.
These features are qualitatively different from those seen in SU($3$).

Furthermore, concerning the existence of the discontinuity, several holographic models also show a clear phase transition 
of the confinement as a function of $l$~\cite{Nishioka:2006gr, Klebanov:2007ws, Witten:1998zw,Sakai:2004cn}.
However, they are relevant to large-$N_c$ Yang-Mills theories only in the far infrared region.

Assuming that the monotonic decrease of the entropic $c$-function results from the confinement,
it is worth comparing the continuous behavior of the entropic $c$-function with another observation of the confinement, 
namely, the static quark-antiquark potential.
The lattice data for the static potential (see {\it e.g.} a review paper~\cite{Greensite:2003bk}) reproduces the Coulomb potential 
of a quark-antiquark pair for short distances, which is seen in the perturbative picture. 
On the other hand, it shows a linear potential for long distances that is a signal of confinement.
For intermediate distances, the lattice data smoothly connect the two regimes.
Our results  are analogous to the behavior of the static potential in the whole regime.

At short distance, the observed value of the entropic $c$-function, Eq.~(\ref{eq:result-Cfn}), can be understood reasonably well in terms of the degrees of freedom of gluons.
First, we note in case of SU($2$) gauge theory, most of the data for $C(l)_{\mbox{SU($2$)}}$ at $l \le 0.3$ fm (Fig.~$6$ in Ref.~\cite{Buividovich:2008kq}) are located in the range of $C=0.07$--$0.08$~\footnote{We thank P.~V.~Buividovich for kindly providing us with the raw data.}.
Scaling down our value for Eq.~(\ref{eq:result-Cfn}) proportional to ($N_c^2-1$), we obtain $C_{\mbox{SU($2$)}} \approx 0.077$.

Calculating the value of the entropic $c$-function for free gluon theory requires an independent quantitative discussion.
The entanglement entropy in four-dimensional free theory is expressed as
\beq
S_A(l) &=& K |\partial A| \left [\frac{1}{a^2} - \frac{1}{l^2} \right].
\eeq
The coefficient, $K \sim 0.0049$~\cite{Ryu:2006ef}, is obtained for the free real scalar theory. 
Assuming that the contribution of a free gauge boson is approximated with two real scalars, and taking into account eight color degrees of freedom in the SU($3$) gauge theory,
we get the following estimate: 
\beq
C(l)_{\mathrm{free}} \sim 0.1568.\label{eq:result-Ryu-Takayanagi}
\eeq
Keeping in mind the approximations made, the prediction Eq.~(\ref{eq:result-Ryu-Takayanagi}) falls remarkably close to our observed value, Eq.~(\ref{eq:result-Cfn}).
The slight discrepancy may be also caused by additional degrees of freedom, such as those due to glueballs, other excited states or topological ground states in the lattice data, while Eq.~(\ref{eq:result-Ryu-Takayanagi}) is obtained in a perturbative vacuum.

We examine the validity of each analysis, in particular, we consider the finite volume effect, 
continuum extrapolation, and replica number dependence.
To estimate the finite volume effect, we carried out a simulation with twice larger lattice extent in each direction with fixed ($\beta, L, \Delta L$).
We found that the finite volume effects are negligible compared to the statistical error.
Next, to estimate the discretization effects, we investigate  the $a$ dependence of the entropic $c$-function.
Clearly, there is no $a$ dependence as shown in Fig.~\ref{fig:C-fn}.
Moreover, we carried out a simulation with a fixed physical lattice extent with half the lattice spacing.
Although the statistical error is still large, the results with the halved spacing are consistent with the results shown in Fig.~\ref{fig:C-fn}.
The replica number dependence was studied by generating the data on $n=3$ replica lattice.
The next-to-leading correction for the $n$-th derivative of the entanglement entropy is smaller than the statistical error of our main results.

In summary, we present the first non-perturbative determination of the entropic $c$-function, which is the $l$-dependence of the entanglement entropy in the SU($3$) pure gauge theory, by using lattice QCD simulations.
We utilize the replica method to obtain the entropic $c$-function, which is consistent with the maximally gauge invariant definition.
At short distances, the entropic $c$-function is proportional to the degree of freedom of gluons, and it vanishes at long distances.
In addition, the change between those two regimes occurs smoothly around a distance that is consistent with the QCD scale.
Several systematic uncertainties are under 
control in our results.

For our future work, we note that it will be straightforward to extend the present study to QCD with dynamical fermions.
Although the exact order parameter for quark confinement is as yet unknown, the entropic $c$-function is expected to provide a new insight for the confinement from the viewpoint of the effective degrees of freedom.
The application of this to finite temperature QCD will also be interesting.
The entropic $c$-function at finite temperature gives the thermal entropy density in the long distance limit.
Preliminary results have already been presented in Ref.~\cite{Nakagawa:2011su}, and as expected, these are roughly consistent with the thermal one obtained by the other approaches~\cite{Boyd:1996bx}--\cite{Asakawa:2013laa}.
Furthermore, 
determining the length of $l$ at which the entanglement entropy density becomes consistent with the thermal entropy density, 
would give a quantum correlation length for the quark-gluon plasma phase.

As an other direction, several recent studies have found various conformal field theories for nonabelian gauge theories 
by using a lattice numerical simulation to observe the non-perturbative running coupling constant ~\cite{Appelquist:2007hu,Itou:2012qn}.
Applying the present method to such conformal systems, the entropic $c$-function tells us the universal quantity related with the central charge for the four dimensional conformal field theories.

\section*{Acknowledgements}

We would like to thank S.~Aoki, P.~Buividovich, J.-W. Chen, M.~Endres, T.~Iritani, K.~Ishikawa, 
S.~Matsuura, S.~Motoki, K.~Nii, M.~Nozaki, T.~Onogi, N.~Shiba, and T.~Takayanagi for useful discussions and comments. 
We also thank the Yukawa Institute for Theoretical Physics (YITP), Kyoto University. 
Discussions during the YITP workshop YITP-T-14-03 on {\it Hadrons and Hadron Interactions in QCD}
 were useful for completing this work.

Numerical simulations were performed on Hitachi SR16000
at YITP, Kyoto University, on a NEC SX-8R and SX-9 at the Research Center for Nuclear Physics (RCNP) Osaka University, and on a Hitachi SR16000 at KEK under its
Large-Scale Simulation Program (Nos.~14/15-05). 
We acknowledge the Japan Lattice Data Grid for data transfer and storage.
The work of K.~N. and A.~N. are supported in part by a Grant-in-Aid for Scientific
Researches~ No. 00586901, 26610072, and 15H03663. 
E.~I. and K.~N. are supported
in part by Strategic Programs for Innovative Research (SPIRE) Field~5. 

The work was completed due to support of the RSF grant  15-12-20008.

\end{document}